%% file: jouralDHN.tex
\documentclass[final,5p,times]{elsarticle}

\usepackage{amssymb}
\usepackage{textcomp}
\usepackage{graphicx}
\usepackage{amsmath}
\usepackage{graphicx}
\usepackage{eurosym}
\usepackage{algorithm}
\usepackage{algorithmic}
\usepackage{url}

\journal{Energy and Buildings}

\begin{document}

\begin{frontmatter}



\title{Model-Free Control of Thermostatically Controlled Loads Connected to a District Heating Network}

\author[label1,label]{Bert J. Claessens\fnref{restorekak}}
\author[label1,label]{D. Vanhoudt}
\author[label1,label]{J. Desmedt}
\author[label2,label]{F. Ruelens}
\fntext[restorekak]{Bert J. Claessens is currently working at REstore and can be contacted at bert.claessens@restore.eu}
\address[label1]{Energy departement of the Research Institute VITO, 2400 Mol, Belgium }
\address[label2]{Department of Electrical Engineering of KU~Leuven, Kasteelpark Arenberg 10, bus 2445, 3001 Leuven, Belgium}
\address[label]{EnergyVille, Thor Park 8310, 3600 Genk, Belgium}

\begin{abstract}
Optimal control of thermostatically controlled loads connected to a district heating network is considered a sequential decision-making problem under uncertainty. The practicality of a direct model-based approach is compromised by two challenges, namely scalability due to the large dimensionality of the problem and the system  identification required to identify an accurate model. To help in mitigating these problems, this paper leverages on recent developments in reinforcement learning in combination with a market-based multi-agent system to obtain a scalable solution that obtains a significant performance improvement in a practical learning time. The control approach is applied on a scenario comprising 100 thermostatically controlled loads connected to a radial district heating network supplied by a central combined heat and power plant. Both for an energy arbitrage and a peak shaving objective, the control approach requires 60 days to obtain a performance within 65\% of a theoretical lower bound on the cost.

\end{abstract}


\end{frontmatter}

\input{introduction}

\input{RelatedWork}

\input{ProblemDescription}

\input{DHNController}

\input{EvaluationDHN}
\input{conclusionsDHN}

\section*{References}
\bibliographystyle{elsarticle-num} 
\bibliography{references}

\end{document}

%% file: introduction.tex
\section{Introduction}
\label{s.Introduction}
A District Heating Network (DHN) offers the opportunity to provide the collective heat demand of a cluster of geographically concentrated buildings through a set of central heat sources. This allows the use of centralized production techniques with an efficiency exceeding that of distributed production. Combined Heat and Power plants (CHPs) are a prominent example, as 80-90\% of the primary energy is converted to heat and electricity \cite{ChristianThesis,Kitapbayev2015823,Sartor2014474}. But also heat from a geothermal source \cite{GDHN} or excess heat resulting from an industrial process can be used as primary heat source.
The heat from the sources is transported through a network of pipes using water as a medium. At each building, heat is extracted in a local substation, resulting in water at a lower temperature, being transported back to the different heat sources.
A typical operational model at the production side is to modulate the power of the heat sources to keep the supply temperature close to a design setting. This basically results in the thermal supply following the thermal demand. 
Heat storage however, can provide demand flexibility enabling flexibility at the production side through demand response approaches. This flexibility allows operational opportunities for cost reduction, examples being peak shaving/valley filling \cite{ClaessensSelfLearning} and energy arbitrage by selling the electricity production of the CHP on the wholesale market \cite{ChristianThesis,Kitapbayev2015823,de2014trading}.
Well referenced embodiments of local heat storage are Thermostatically Controlled Loads (TCLs) \cite{KochThesis} such as a hot water storage tank \cite{Vanthournout} where the heat is stored directly in the water, but also the building envelope \cite{ChristianThesis,KochThesis,Verbeeck,Kensby2015773} can be used to store heat.\\
\indent From an operational point of view, controlling a cluster of TCLs connected to a DHN can be considered as a sequential decision-making problem under uncertainty.
One well studied control paradigm for operational management of a DHN is that of Model Predictive Control (MPC) \cite{SandouMPCDHN,Široký20113079}. When projected on the setting of TCLs connected to a DHN, this requires defining control actions for the central sources as well as for all individual TCLs.
Developing a practical implementation requires one to tackle the problem of \textit{scalability}, as the state dimensionality and number of control variables quickly result in an intractable optimization problem. This is complicated further by non-linear system dynamics.
A second important challenge is that of \textit{system identification} \cite{mathieu2013energy} as identifying an accurate model of both the DHN and all TCLs requires significant amounts of not readily available data and expert knowledge.\\
\indent This work contributes in mitigating operational control challenges for TCLs connected to a DHN by working on these two problems.\\
\indent Scalability: To obtain scalability, a heuristic dispatch approach as described in \cite{Stijn}, is applied to the setting of TLCs connected to a DHN. Instead of calculating an individual control action for each TCL, this approach calculates a collective control action for the entire cluster of TCLs. A market-based dispatch algorithm is used to translate the collective control action into individual control actions.\\
\indent System identification: Driven by recent developments in Batch Reinforcement Learning (BRL) \cite{ReinforcementLearning,atariRL}, a \textit{blind} model-free approach is considered. As a BRL technique needs no prior information on the system dynamics, this strongly relaxes the system identification requirements, at the cost of a \textit{learning time} and sub-optimal performance.\\
\indent In Section \ref{s.RelatedWork}, an overview is given on related research regarding the control of large clusters of TLCs and model based controllers for DHN. In Section \ref{s.Approach}, the decision-making problem is formalized as a Markov Decision Problem (MDP).
In Section \ref{s.TSADHN}, the control approach as used in this work, is described in detail.
In Section \ref{s.Evaluation}, an evaluation of the controller performance is provided based upon a simulation scenario comprising 100 TCLs connected to a DHN. The simulation scenario is sufficiently complex to evaluate the contributions of the control approach, but also comprehensibly enough to allow for an analysis which is not obfuscated by complexity of the scenario. 
Finally in Section \ref{s.conclusionsDHN}, the conclusions are provided, as is a discussion on the results.

%% file: RelatedWork.tex
\section{Related Work}
\label{s.RelatedWork}
In this section a non-exhaustive overview is given of related work regarding both the control of large clusters of TCLs and model-based control applied to a DHN.
\subsection{Controlling a cluster of TCLs}
The \textit{curse of dimensionality} \cite{Bertsekas} lurks around the corner when managing the flexibility present in a large cluster of TCLs. This is attributed to the dimensionality of the state space and the large amount of control variables.
To this end, significant recent work \cite{mathieu2015arbitraging,tracers, georges2016direct,BiegelHP,Hu2015229} has focused on providing computationally tractable solutions for large clusters of TCLs. 
In \cite{Mathieu}, a cluster of flexibility carriers represented by generic tank models is considered with the objective of providing day-ahead modulation services to a transmission system operator. Even though all models are linear, a formal branch and bound based optimization approach quickly becomes intractable. The main contribution of the work is a heuristic method including a state dependent dispatch algorithm in combination with an iterated local search technique. High quality solutions to a test problem are presented, obtained within a practical calculation time.
The results of a related approach applied in an actual field test comprising 54 heat pumps has been presented  in \cite{BiegelHP}. 
Here an aggregated model of reduced dimensionality was used to determine power set points for the entire cluster in a MPC approach. A heuristic dispatch algorithm was used to convert the aggregated set points to local control actions in the portfolio. In \cite{Hu2015229} a data-driven decision framework has been developed using a meta-heuristic optimization technique.\\
An approach from the same \textit{solution class} is presented in \cite{mathieu2013energy}.
Here a problem of energy arbitrage with a large cluster of TCLs is presented.
An aggregated system model is used in the form of a state bin transition model \cite{Koch}. All TCLs are clustered, based upon their position within their dead-band, resulting in a state vector containing the fraction of TCLs in each state bin. A linear state bin transition model describes the dynamics of this state vector, the dimensionality of which is independent of the number of TCLs in the cluster. This model is used in an MPC resulting in a control action for each state bin. A simple heuristic is used to dispatch the control signals to individual control actions at device level. Although a simplified first order TCL model has been used, the results presented in \cite{mathieu2013energy} show that careful system identification is required. Moreover, in \cite{Zhang} it was argued that a first order TCL model is found lacking, further complicating system identification.\\
A different approach is that of distributed optimization \cite{Gatsis,Bosman}, where the centralized optimization problem is decomposed over distributed agents who interact through \textit{virtual} prices. For example in \cite{BiegelDD}, distributed MPC through dual decomposition was presented as a means for energy arbitrage of a large cluster of TCLs subjected to a coupling constraint related to an infrastructure limitation. Although mathematical performance guarantees can be provided under sufficient assumptions, the method heavily relies on the accuracy of local models and has stringent communication and computation requirements due to its iterative character. 
An example of how an MPC controller can be used at building level is detailed in \cite{Široký20113079}.

\subsection{Model-Based DHN Control} 
When implementing a model-based control approach for a DHN, be it centralized or distributed, one is confronted with (1) the non-linearities in the dynamics of a DHN, and (2) the slow time scales compared to e.g. an electric network \cite{Dirk}. Taking these effects into account is essential for a good performance of the controller. Several model-based optimization approaches have been identified in literature, explicitly incorporating the dynamics of the DHN. For example in \cite{SandouMPCDHN} a simplified model has been derived that is used together with sequential quadratic programming. In \cite{Ikonen} approximate dynamic programming \cite{ADP} has been used taking advantage of permutational symmetries of the DHN dynamics. A model-based approach using fuzzy direct matrix control to mitigate non-linearities of the DHN dynamics can be found in \cite{GrosswindFDMC}. 
Although model-based solutions can have excellent performance, accurate models of the DHN and the consumers coupled to the DHN are required, tuning and shaping these models is considered an \textit{expert task} making a generic roll-out of this technology challenging \cite{Pinson2009163}.\\
\indent A scalable model-free solution solution is presented in \cite{Booij}, here a market-based multi-agent system is used to match thermal and electric demand and supply. Although this approach is scalable, it does not take into account the DHN dynamics and follows a myopic control strategy.
An approach combining an auction-based multi-agent system with a central optimization, taking into account a forecast of the total heat demand can be found in \cite{ChristianThesis}.

%% file: ProblemDescription.tex
\section{Problem description}
\label{s.Approach}
Inspired by Gemine \textit{et al.} \cite{gemine}, this section presents a problem formulation of the sequential decision-making problem related to optimal control of a cluster of TCLs connected to a DHN. 
In a second step, the control problem is cast onto a Markov Decision Process (MDP).  
\begin{figure}[h!]
\centerline{\includegraphics[width=0.95\columnwidth]{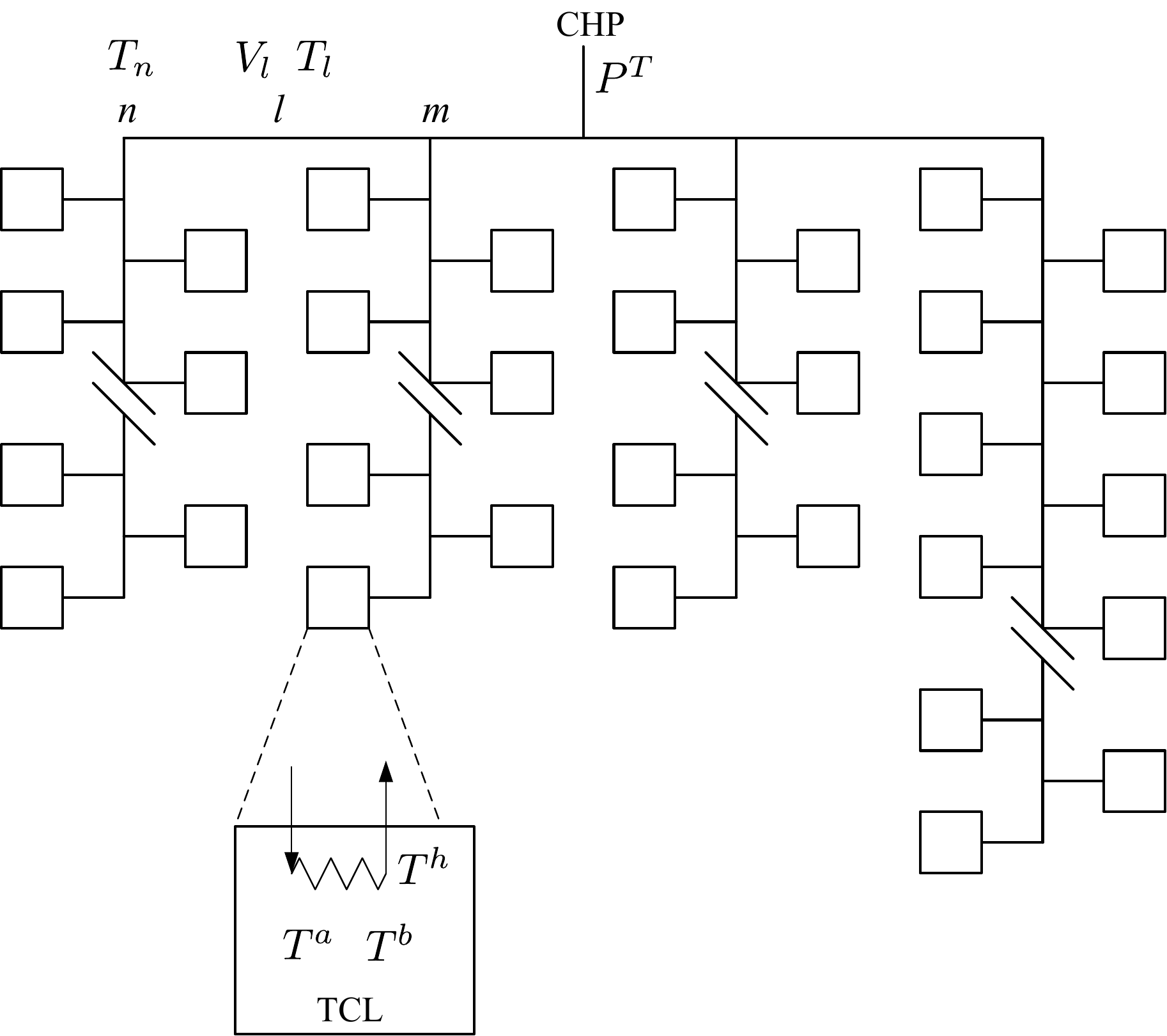}}
\caption{Illustration of the district heating network configuration as used in this work. A set of 100 Thermostatically Controlled Loads (TCLs) is connected to a district heating network, the thermal energy is produced by one central Combined Heat and Power (CHP) plant at the top.}
\label{fig.OverViewDHN}
\end{figure}
\subsection{Test scenario network}
To make the problem formalism more tangible, references are made to the test scenario as illustrated in Figure \ref{fig.OverViewDHN} and detailed in Section \ref{s.Evaluation}. The test scenario comprises 100 TCLs connected to a DHN. One central CHP is assumed to provide the heat to the buildings through a DHN. 
\subsection{Problem components}
\subsubsection{DHN}
A DHN contains a set of nodes and pipes connecting these nodes. Each pipe $l \in \mathcal{L}$ connects two nodes $n,m \in \mathcal{N}$ and is characterized by its length $L_{l}$, diameter $D_{l}$ and heat loss coefficient $\alpha_{l}$. The flow speed of the medium at time $t\in \mathcal{T}$ in pipe $l$ is denoted by $V_{l,t}$ and its average temperature by $T_{l,t}$. The temperature in each node $n$ at time $t$ is characterized by $T_{n,t}$.
\subsubsection{TCLs and heat production}
The set $\mathcal{D}$ contains the TCLs connected to the DHN, each TCL is assumed connected to a node. For modeling purposes a set of relevant temperatures is associated to each TCL $d \in \mathcal{D}$ at time $t$, i.e. the air temperature $T_{d,t}^{a}$ as measured by a local thermostat, $T_{d,t}^{b}$ a temperature corresponding to a building envelope \cite{Verbeeck} and $T_{d,t}^{h}$ the heating system return temperature.
The control available at the level of a TCL is to decide whether or not to extract heat from the DHN, corresponding to a binary value $b_{d,t} \in \{0,1\}$. The thermal inertia of the DHN and TCLs is used for storage, no separate hot water storage is considered.\\
\indent Besides TCLs also production units $p \in \mathcal{P}$ are connected to the DHN in specific nodes. As illustrated in Figure \ref{fig.OverViewDHN}, here a single CHP is used as heat source. At every time step $t$, an input power $P^{I}_{p,t}$, a thermal output power $P^{T}_{p,t}$ and an electric output power are $P^{e}_{p,t}$  are associated to $p$ in a node $n$.
The relationship between these powers is defined as:
\begin{equation}
P^{T}_{p,t}+\varphi P^{e}_{p,t} = \eta P^{I}_{p,t},~\forall(p,t)\in \mathcal{P} \times \mathcal{T},
\label{eq.CHP}
\end{equation}
with $\eta$ the total fuel utilization ratio of the CHP and $\varphi$ defines the heat to power ratio of the CHP.
In this work $P^{T}_{p,t} \in \left [ 0, P_{p,max}^{T} \right ]$ is considered the control variable.

\subsubsection{Operational limits}
The main interest of the DHN system is to supply a heat service towards the TCLs meeting comfort constraints:
\begin{equation}
\forall (d,t) \in \mathcal{D} \times \mathcal{T}: \underline{T}^{a}_{d} \leq T_{d,t}^{a} \leq \overline{T}^{a}_{d}. 
\label{eq.constraintsTCL}
\end{equation} Here $\underline{\cdot}$ and $\overline{\cdot}$ indicate the lower and upper bound respectively. Common for buildings is to have a constraint on the operative temperature. As a simplification the comfort constraints are here directly related to the air temperature $T_{t}^{a}$.
Besides constraints for the TCLs also constraints for the DHN are relevant, i.e. 
\begin{align}
\forall (n,t) \in \mathcal{N} \times \mathcal{T}:& \underline{T}_{n} \leq T_{n,t} \leq \overline{T}_{n}\\ 
\forall (l,t) \in \mathcal{L} \times \mathcal{T}:& \underline{T}_{l} \leq T_{l,t} \leq \overline{T}_{l},\\
& \underline{V}_{l} \leq V_{l,t} \leq \overline{V}_{l}.\nonumber 
\label{eq.constraintsDHN}
\end{align}
\subsection{Sequential decision making}
Driven by the possibility of using techniques from Reinforcement Learning (RL) \cite{ReinforcementLearning}, the sequential decision-making process is formulated as a Markov Decision Process (MDP) \cite{Bertsekas,FonteneauAT}. The sequential nature results from inter-temporal constraints related to the dynamics of the DHN and the TCLs. Decisions made at time step $t$ impact possible actions allowed at future states.  
An MDP is defined by its state space $X$, its action space $U$, and a transition function $f$:
\begin{equation}
\mathbf{x}_{t+1}=f(\mathbf{x}_{t},\mathbf{u}_{t},\mathbf{w}_{t}),
\end{equation} describing the dynamics from state $\mathbf{x}_{t}\in X$ to $\mathbf{x}_{t+1}$, following the control actions $\mathbf{u}_{t} \in U$ subject to a random process, $\mathbf{w}_{t} \in W$, where $\mathbf{w}_{t}$ is drawn from a probability distribution $p_{w}(\cdot,\mathbf{x}_{t})$. 
Each transition is accompanied by a cost signal $c_{t}$:
\begin{equation}
c_{t}(\mathbf{x_{t}},\mathbf{u_{t}},\mathbf{x_{t+1}})=\rho(\mathbf{x}_{t},\mathbf{u}_{t},\mathbf{w}_{t}),
\end{equation}   
with $\rho$ the cost function. 
\subsubsection{State description}
\label{s.stateDes}
Following the notation provided in~\cite{RuelensBRLDevice}, the state of the system is assumed to be spanned by time dependent state information $X_{t}$, controllable state information $X_{phys}$ and uncontrollable exogenous state information $X_{ex}$ \cite{Bertsekas}. 
The time dependent state information $X_{t}$ describes the time information relevant for the dynamics, e.g. the quarter of an hour in the day or the day in the week. In this work 
$X_{t} = \left\lbrace 1,...,96 \right\rbrace $. 
The controllable state information $X_{phys}$ represents the state of the DHN and the TCLs:
\begin{align}
\mathbf{x_{phys,t}}= (& T_{1,t}^{a},T_{1,t}^{b},T_{1,t}^{h},...,T_{|\mathcal{D}|,t}^{a},T_{|\mathcal{D}|,t}^{b},T_{|\mathcal{D}|,t}^{h}, \\
&T_{1,t},...,T_{|\mathcal{N}|,t},T_{1,t},V_{1,t},...,T_{|\mathcal{L}|,t},V_{|\mathcal{L}|,t}).
\end{align}
The uncontrollable exogeneous state information comprises the physical parameters relevant for the dynamics of the system that can not be influenced by $u_{t}$. Examples being the outside temperature $T_{out}$, solar irradiation $S$, wind speed and direction $\overrightarrow{v}$ \footnote{The outside temperature, solar irradiation and wind information are assumed constant over the DHN and the buildings.} and local electric consumption $e_{d}$.
\begin{align}
\mathbf{x_{ex,t}}=\left(T_{out,t},S_{t},\overrightarrow{v}_{t},e_{1,t},...,e_{|\mathcal{D}|,t}\right).
\end{align}
It is $\mathbf{x_{ex,t}}$ that represents $\mathbf{w_{t}}$. If there is no correlation between $\mathbf{w_{t}}$ and $\mathbf{w_{t+1}}$ ($p_{w}(.)$), $\mathbf{x_{ex,t}}$ can be omitted from the state information \cite{Bertsekas} in the MDP, however by having $\mathbf{x_{ex,t}}$ in the state vector a first order correlation is assumed\footnote{This can readily be extended to include information several time steps back.} ($p_{w}(.|\mathbf{x_{ex}})$). 

\subsubsection{Control actions}
The control vector includes the control actions of the TCLs and the thermal output power of the central CHP:
\begin{equation}
\mathbf{u_{t}}=(P_{p,t}^{T},b_{1,t},...,b_{|\mathcal{D}|,t}).
\end{equation}  
\subsubsection{DHN Dynamics}
To model the dynamics of the DHN (as used in the evaluation),  a quasi-dynamic approach is followed \cite{Dirk} as pressure and flow change orders of magnitude faster than the temperature of the water. In a first step, a hydraulic simulation is performed, in a second step the thermal dynamics are calculated. For the hydraulic calculations the approach as proposed in \cite{Valdimarsson1993} is followed. This results in applying Kirchhoff's laws, with the consideration that there exists a non-linear relationship between pressure and flow rate.\\ 
\indent To calculate the thermal dynamics, the node model as presented by Benonysson \cite{Benonysson} has been used, essentially solving the following equation for every pipe section:
\begin{equation}
c_{pw}\frac{\partial mT_{w}}{\partial t}+c_{pw}\frac{\partial \dot{m}T_{w}}{\partial x}+hA(T_{w}-T_{g})=Q.
%
%
\label{eq.3}
\end{equation} 
With $m$ the mass of the water, $c_{pw}$ the thermal capacity of water, $T_{w}$ the water temperature, $Q$ the heat demand  and $h$ the heat transfer coefficient between the water and the ground.  The surface of the pipe considered is $A$ and $T_{g}$ the local ground temperature.  
\subsubsection{TCL Dynamics}                       
For the building models, a lumped capacitance model is used i.e. an electric analogue following Verbeek \cite{Verbeeck}. The model includes the temperature dynamics of the inside air, a building envelope and the heating system return temperature \cite{Dirk}. Besides heat losses to the ambient air, also wind speed dependent air infiltration losses are included, as are uncontrolled heating due to solar irradiation and local electric consumption \cite{Dirk}.

\subsubsection{Cost signal}
Finally also a cost function $c$ needs to be defined: $c: X \times U \times X \rightarrow \mathbb{R}$, in this work, two objectives are regarded, i.e. energy arbitrage $c_{\lambda}$, responding to an external price $\mathbf{\lambda}$ and peak shaving/valley filling $c_{ps}$.
The cost functions are defined as:
\begin{equation}
c_{\lambda}(\mathbf{x_{t}},\mathbf{u_{t}},\mathbf{x_{t+1}})=P_{p,t}^{T}\lambda_{t}\Delta t.
\label{eq.rewardarbitrage}
\end{equation} with $\lambda_{t}$ the effective price for producing thermal energy at time step $t$.
For the peak shaving objective, the cost is expressed on a daily basis:
\begin{equation}
c_{ps}(\mathbf{x_{1}},\mathbf{u_{1}},...,\mathbf{u_{T}})=\mathbb{E}\left\{\left\|P_{p,1}^{T},...,P_{p,T}^{T}\right\|_{\infty}\right\}.
\label{eq.rewardpeakshaving}
\end{equation}
The objective of this work is to find a control policy $\pi: X \rightarrow U$ that will minimize the $T$-stage return $J^{\pi}(x_{1})$ starting from state $x_{1}$ defined as:
\begin{equation}
J^{\pi}(\mathbf{x_{1}}) = \mathbb{E}\left(R^{\pi}(\mathbf{x_{1}},\mathbf{w_{1}},...,\mathbf{w_{T}})\right). 
\label{eq.J}
\end{equation}
with 
\begin{equation}
R^{\pi}(\mathbf{x_{1}},\mathbf{w_{1}},...,\mathbf{w_{T}}) = \sum_{t=1}^{T}{\rho(\mathbf{x_{t}},\pi(\mathbf{x_{t}}),\mathbf{w_{t}})} 
\label{eq.rewardf}
\end{equation} from the understanding that an optimal policy $\pi^{*}$ satisfies the Bellman equation:
\begin{equation}
J^{\pi^{*}}(\mathbf{x})=\underset{\mathbf{u}}{\mathrm{min}}~\underset{\mathbf{w}\sim P_{w}(.|\mathbf{w})}{\mathbb{E}}\lbrace \rho(\mathbf{x},\pi(\mathbf{x}),\mathbf{w})+J^{\pi^{*}}(f(\mathbf{x},\pi(\mathbf{x}),\mathbf{w})) \rbrace
\label{eq.Bellman}
\end{equation}
When an accurate model is available, typical techniques to find near-optimal policies in an MDP framework are value iteration, policy iteration, direct policy search \cite{Busoniu} and tree search algorithms such as optimistic planning \cite{BusoniuOptimistic}. In \cite{Ikonen} for example, an approximated value iteration approach is followed to determine a control policy for the control of a DHN.\\
In this work, a model-free approach is explored. Driven by promising results \cite{Ernst, atariRL, FonteneauAT}, Batch Mode Reinforcement Learning (BRL) techniques are investigated, detailed in Section \ref{s.TSADHN}.

%% file: DHNController.tex
\section{DHN controller approach}
\label{s.TSADHN}
This section describes a \textit{pragmatic} control approach building upon recent results in BRL and market-based multi-agent systems.
\begin{figure}[t!]
\centerline{\includegraphics[width=1.0\columnwidth]{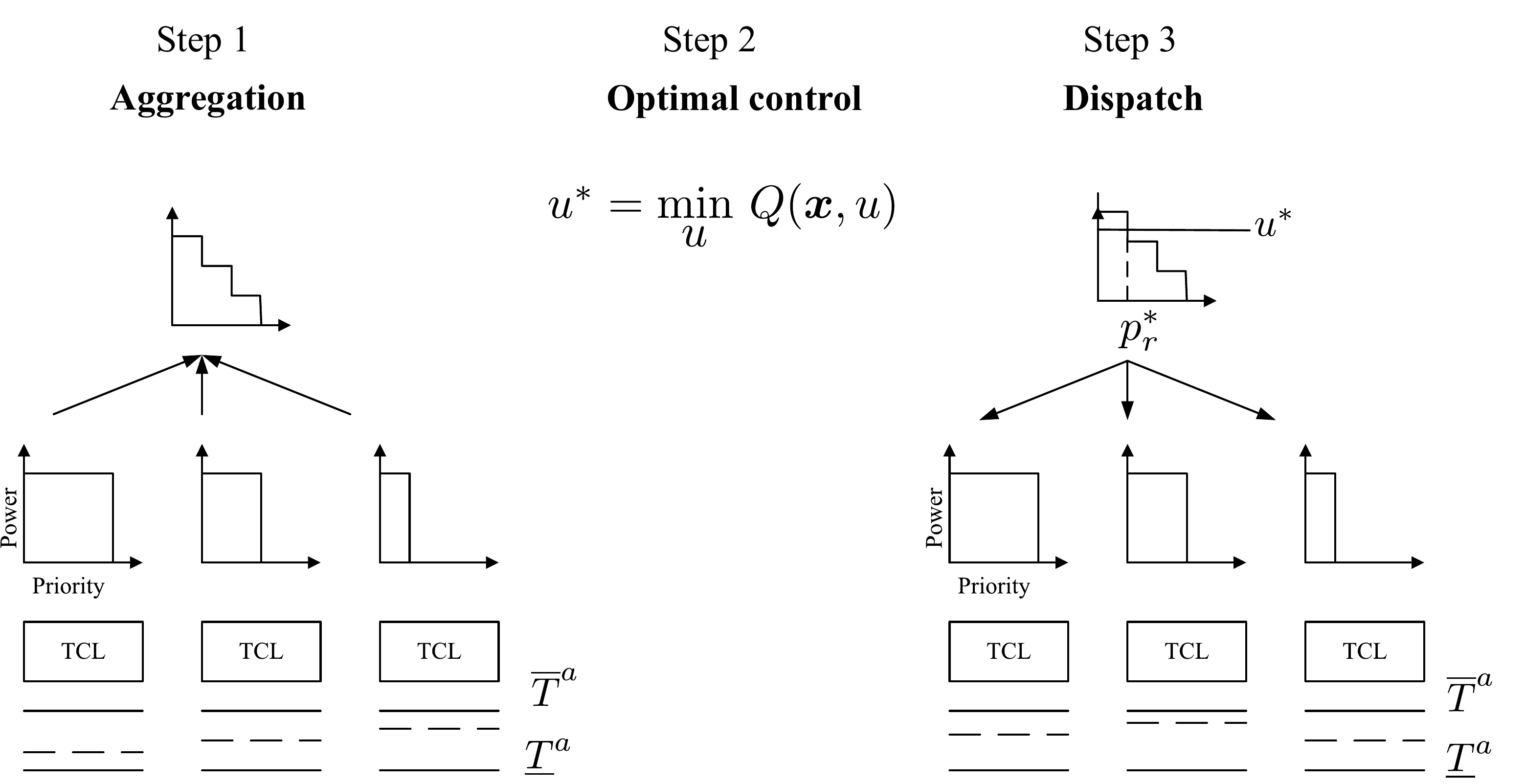}}
\caption{Overview of the three-step approach. In a first (left) aggregation step the flexibility and state information of the Thermostatically Controlled Loads (TCLs) are aggregated. In a second optimization step (middle), the optimal control action ${u^{*}}$ for the entire cluster is determined, finally in step three (right) ${u^{*}}$ is dispatched using a market-based multi-agent system.}
\label{fig.OverViewTSA}
\end{figure}
The control approach illustrated in Figure \ref{fig.OverViewTSA} and followed in this work is based upon a Three Step Approach (TSA) as presented in \cite{Stijn,Arnout} following a similar strategy as followed in \cite{BiegelHP,Koch}.
In a first step, (1) all relevant (and practically available) state information is collected, e.g. the temperature information from the TCLs. From this information a limited set of features \cite{Bertsekas} is extracted, resulting in a low-dimensional representation of the system state.
In a second step (2) a control action for the entire cluster of TCLs is extracted from a policy $\pi^{*}$ determined offline on given time intervals. In a third and last step (3), this control action is dispatched over the different TCLs using a market-based multi-agent system. This process is repeated following a receding horizon approach. In the following, a more detailed description of the three steps is presented.

\subsection{Step 1: Aggregation}
\label{s.stepOne}
In the first step, state information as described in Section \ref{s.stateDes} is retrieved from the system. From a practical perspective however, not all state information is readily available. At building level, only the air temperatures as measured by a local thermostat are assumed available, i.e. the air temperatures measured in the buildings, $T_{d,t}^{a}, \forall d \in \mathcal{D}$. Furthermore measurements of the outside air temperature $T_{out,t}$ are assumed available, as is the water temperature from a subset of nodes in the supply side of the DHN, i.e. $T_{n,t}, \forall n \in \mathcal{N}_{s} \subset \mathcal{N}$ and the return side\footnote{Only temperatures at a limited set of nodes are assumed available.}  $T_{n,t}, \forall n \in \mathcal{N}_{r} \subset \mathcal{N}$.
This information is further aggregated, formally this can be seen as a feature extraction \cite{Bertsekas}, which reduces the dimensionality of the decision-making process. In this work, the feature extraction is \textit{handcrafted}, resulting in the following effective state vector $x_{t}$.
\begin{align}
x_{t} &= (\left\langle T_{t}^{a}\right\rangle,T_{out,t},T_{s},T_{r})\label{eq.state}\\
\left\langle T_{t}^{a}\right\rangle& = \frac{\sum_{d \in \mathcal{D}}{T_{d,t}^{a}}}{|\mathcal{D}|}\\
T_{s}& = \frac{\sum_{n \in \mathcal{N}_{s}}{T_{n,t}}}{|\mathcal{N}_{s}|}\\
T_{r}& = \frac{\sum_{n \in \mathcal{N}_{r}}{T_{n,t}}}{|\mathcal{N}_{r}|}
\label{eq.statevector}
\end{align}
Although, more generic dimension reduction techniques such as autoencoders can be used \cite{RiedmillerAuto}, the aim of this work is to understand what performance can be obtained starting from this limited state description. 
Alternatively, a convolutional neural network as presented by the the authors in~\cite{DDRBert} could be used to automatically extract relevant state-time features, allowing to add historic observations to the state~\cite{atariRL, Bertsekas}. \\
\indent To facilitate the dispatch step as explained in Section  \ref{s.realtime}, a bid-function is defined for every TCL \cite{ClaessensSelfLearning,Stijn,KlaasEventBased}. In \cite{ClaessensSelfLearning}, 
the bid function of a device is expressed as the electric power consumed versus a heuristic ($p_{r}$). Above a corner value $p_{c,d}$ the bid function is zero:
\begin{equation}
p_{c,d}= 1-SoC=\frac{\overline{T}^{a}_{d}-T_{d,t}^{a}}{\overline{T}^{a}_{d}-\underline{T}^{a}_{d}}.
\label{eq.3}
\end{equation}
Determining this heuristic is considered relatively straightforward as it requires only the air temperature as measured by the thermostat and the upper and lower temperature bound. Defining the thermal power extracted from the DHN by a TCL when switched on, is less so \cite{Dirk}. To relax this 
requirement we assume an estimate of the flowrate ($L_{d}$) when switched on is available. 
The flowrate $L_{d}$ is defined as follows. First the set temperature of the indoor heat supply system, which is a function of the outdoor temperature, is calculated. Then, a model for the substation heat exchanger is used to determine the flow rate extracted from the DHN, by which the outlet temperature of the heat exchanger meets the set temperature.
 Using this value instead of the actual power results in the following bid function for building $d$:
\begin{equation}
b_{d}(p_{r})=L_{d}(1-H(p_{r}+SoC-1)),
\label{eq.bidBoiler}
\end{equation} here $H$ corresponds to the heaviside function. 

\begin{algorithm}
\caption{Overview fitted Q-iteration}
\label{TSASL}
\begin{algorithmic}[1] 
\algsetup{linenosize=\tiny}
\renewcommand{\algorithmicrequire}{\textbf{Input:}}
\REQUIRE $\mathcal{F}_{D} =\left\{(x_{l_{s}},u_{l_{s}},c_{l_{s}},x_{l_{s}}'),l_{s}=1,...,n_{s}\right\}$, regression algorithm \cite{Geurts}.\\
\STATE let $\hat{Q}_{0}$ be zero everywhere on $X$ $\times$ $U$
\REPEAT
\FOR {$l_{s}=1,...,n_{s}$}
\STATE ~~$Q_{l+1,l_{s}}'\leftarrow r_{l_{s}}+\text{min}_{u'}\hat{Q_{l}}(x_{l_{s}}',u')$ 
\ENDFOR
\STATE use regression to obtain $\hat{Q}_{l+1}$ from $\left\{\left((x_{l_{s}},u_{l_{s}}),Q_{l+1,l_{s}}'\right),l_{s}=1,...,n_{s}\right\}$
\UNTIL{$\hat{Q}_{l+1}$ is satisfactory}
\RETURN $\hat{Q}=\hat{Q}_{l+1}$
\end{algorithmic}
\end{algorithm}

\subsection{Step 2: Batch Reinforcement Learning}
\label{s.BRL}
In the second step, a control action $u_{t}$ is selected once every 15 minutes, following the policy $\pi$. The control action $u_{t}$ is selected for the entire cluster which is projected onto individual control actions as described in Section  \ref{s.realtime}.
One of the main goals of this work is to explore to what extent (model-free) reinforcement learning can be used to determine $\pi$. Reinforcement Learning (RL) is a model-free control approach that learns a policy by interaction with the system \cite{Busoniu}. In recent literature, RL (mainly in the form of Q-learning \cite{NeilRL,karaRL,Mocanu2016646}) has been presented as an effective model-free learning approach for DR applications. The practicality however, suffers from slow convergence \cite{ReinforcementLearning,Busoniu} and the \textit{curse of dimensionality} \cite{Bertsekas}. These challenges can be partially mitigated by using past interactions and appropriate function approximators in a BRL strategy \cite{Busoniu,Xu20141}. A popular BRL approach is that of Fitted Q-Iteration (FQI) introduced by Ernst \textit{et al.} in \cite{FQI}, especially in combination with extremely randomized trees as regression technique \cite{Geurts}. In \cite{Ernst} the authors conclude that especially for non-linear control problems, FQI can be a valuable alternative to MPC approaches with the extra advantage that FQI is a \textit{blind} technique. Moreover, FQI and MPC can strengthen each other \cite{Ernst,MABRL}. Although several BRL techniques have been proposed in the literature \cite{FonteneauAT,FQI,RiedmillerAuto}, this work focuses on FQI using extremely randomized trees as regression algorithm \cite{Geurts}. Comparison to the performance of other BRL approaches is considered outside the scope of this work.\\  
\indent Following \cite{FQI}, an approximation of the state-action value function ${Q}(x_{t},u_{t})$,  $\hat{Q}(x_{t},u_{t})$ is built on a daily basis\footnote{This is done for practicality, as the simulations cover a time span of several months, in a real-life application, $\hat{Q}$ should be constructed more frequently, following Algorithm 1.} from a batch of four-tuples $\mathcal{F}_{D}$:
\begin{equation}
\mathcal{F}_{D} =\left\{(x_{l_{s}},u_{l_{s}},c_{l_{s}},x_{l_{s}}'),l_{s}=1,...,n_{s}\right\}.
\label{eq.bidBoiler}
\end{equation}
With the state vector as defined in equation (\ref{eq.state}).
Algorithm 1 is used to obtain $\hat{Q}(x_{t},u_{t})$.
During the day, the control action $u_{t}$ is selected with a probability defined by \cite{Powell}:
\begin{equation}
P(u_{t}|x_{t}) = \frac{e^{\hat{Q}(x_{t},u_{t})/\tau_{D}}}{\sum_{\tilde{u}}{e^{\hat{Q}(x_{t},\tilde{u})/\tau_{D}}}}.
\label{eq.Boltzmann}
\end{equation}
The temperature $\tau_{D}$ is decreased on a daily basis according to a harmonic sequence \cite{ADP}, a high temperature results in more exploration whilst $\tau_{D}\rightarrow 0$ results in a greedy approach:
\begin{equation}
u_{t} =   \underset{u}{\mathrm{arg~min}}~ \hat{Q}(x_{t},u).
\label{eq.greedy}
\end{equation}
In the peak shaving scenario, a control action is determined on a daily basis, defining the average power to be followed for the next day as detailed in \cite{ClaessensSelfLearning}.

\subsection{Step 3: Real-time control}
\label{s.realtime}
\begin{figure}[h!]
\centerline{\includegraphics[width=0.9\columnwidth]{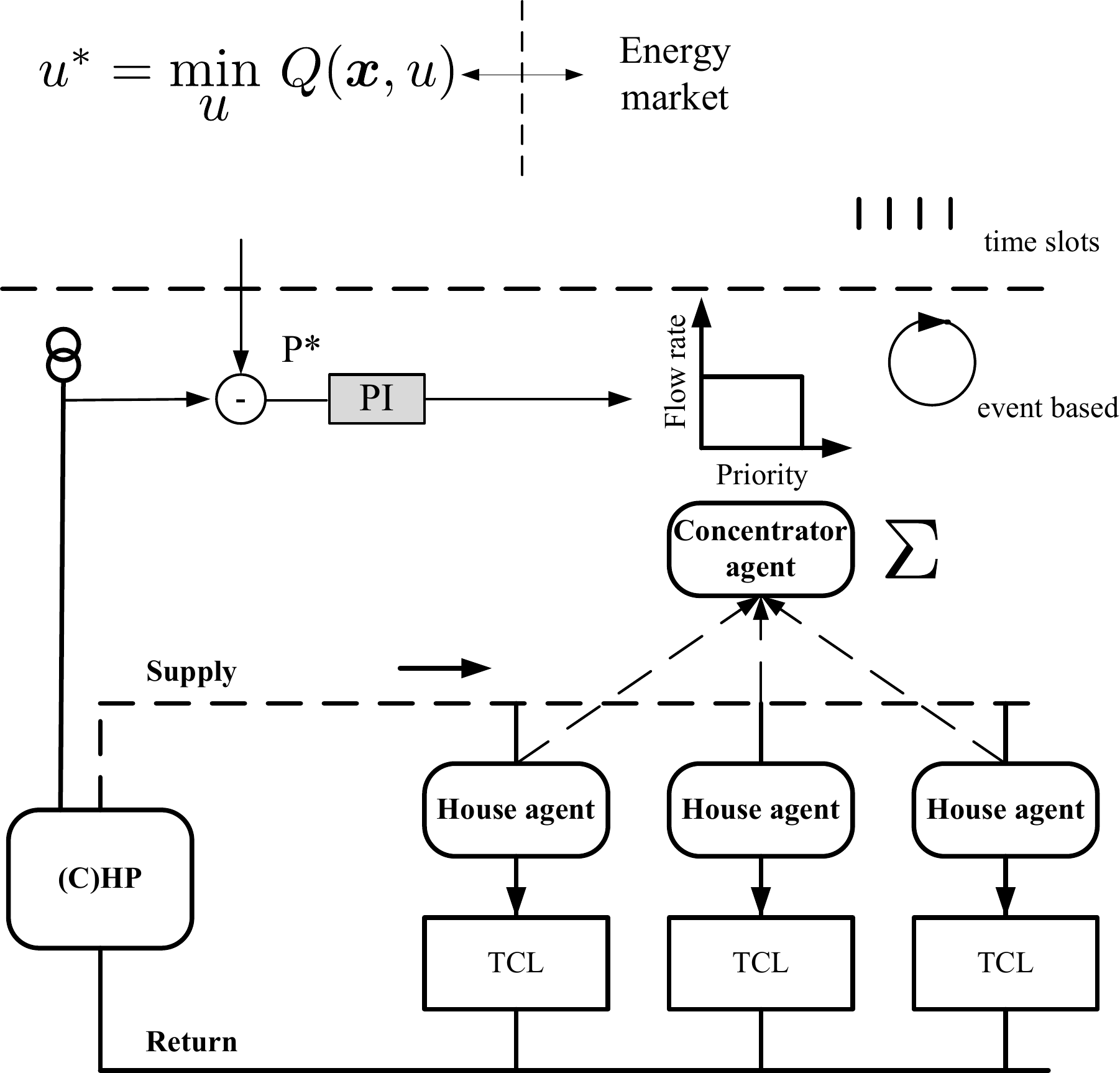}}
\caption{Overview of the controller approach as developed in this work.}
\label{fig.OverViewPV}
\end{figure}

In the third step, the energy corresponding to $u_{t}^{*}$ is dispatched over the cluster of TCLs, using a market-based multi-agent system \cite{Stijn,Hommelberg}. Compared to the work of \cite{ClaessensSelfLearning}, there is a significant difference as only the expected flow rate for each TCL is assumed available.
To this end, a Proportional Integrator (PI) controller (at a central level) managing the flow rates at the different buildings is used. Since hydraulic effects occur nearly instantaneous in a DHN, this will have a direct effect on the flow rate at the source, influencing the power at the source side as the supply setpoint is assumed to be constant in the simulations \cite{Dirk}. 
An overview of the real-time control can be seen in Figure~\ref{fig.OverViewPV}.
As described in Section \ref{s.stepOne}, every TCL is represented by a bid function $b_{d}$. 
After a clearing process (\ref{eq.clearing}), a clearing priority ${p_{r}}^{*}$ is sent back to the different devices:
\begin{equation}
p_{r}^{*} ~= ~\underset{p_{r}}{\mathrm{arg~min}} \left|\sum_{d=1}^{|\mathcal{D}|}{b_{d}(p_{r}})-u\right|,
\label{eq.clearing}
\end{equation}
The devices open or close their local valve according to $b_{d}(p_{r}^{*})$. 

%% file: EvaluationDHN.tex
\section{Evaluation}
\label{s.Evaluation}
To evaluate the performance of the controller described in Section \ref{s.TSADHN}, a set of simulations have been performed. In this section, first a condensed description of the simulation scenario will be presented after which the \textit{tracking} performance of the controller is evaluated. The performance of the controller is evaluated for two distinct objectives, i.e. that of energy arbitrage on a day-ahead energy market and peak shaving/valley filling.
\subsection{Simulation scenario}
As mentioned in Section \ref{s.Introduction}, the scenario is designed to be sufficiently demanding, but also simple enough to allow for an analysis which is not obfuscated by the complexity of the scenario. 
To this end, the (arbitrary) topology as depicted in Figure \ref{fig.OverViewDHN} has been used, i.e. a central CHP ($\eta = 0.97, \phi=0.93, P_{p,max}^{T}=1100$kW) provides heat to 100 TCLs connected to a radial DHN. Each building is located in one of four streets. The total length of the grid is 2.1 km, with pipe diameters ranging from DN25 to DN100. A detailed description of the simulation scenario can be found in \cite{Dirk}.
The TCL building models are lumped capacitance models using an electric analogon comprising capacitances and resistors. The capacitance values are related to the temperature of the inside air, a building envelope and the heating system return temperature \cite{Verbeeck,Dirk}. All 100 buildings models included in the simulation are derived from the same model. Different model parameters are used for each building, by sampling capacitance and resistance values from a normal distribution with a standard deviation of 20\% of the standard value. The standard values correspond to a detached house with a living area of 103 m$^{2}$ and a protected volume of 452 m$^3$. The maximum standard power demand of the building is 9.8 kW at an internal temperature of 20$^{\circ}\mathrm{C}$ and an ambient temperature of -8$^{\circ}\mathrm{C}$. Besides heat losses to the ambient air, also wind speed dependent air infiltration losses are included as is uncontrolled heating due to solar irradiation and local electric consumption \cite{Dirk}. For simplicity the temperature constraints are set the same for all buildings at 19.5 and 20.5$^{\circ}\mathrm{C}$.

\subsection{Tracking performance}
\begin{figure}[h!]
\centerline{\includegraphics[width=1\columnwidth]{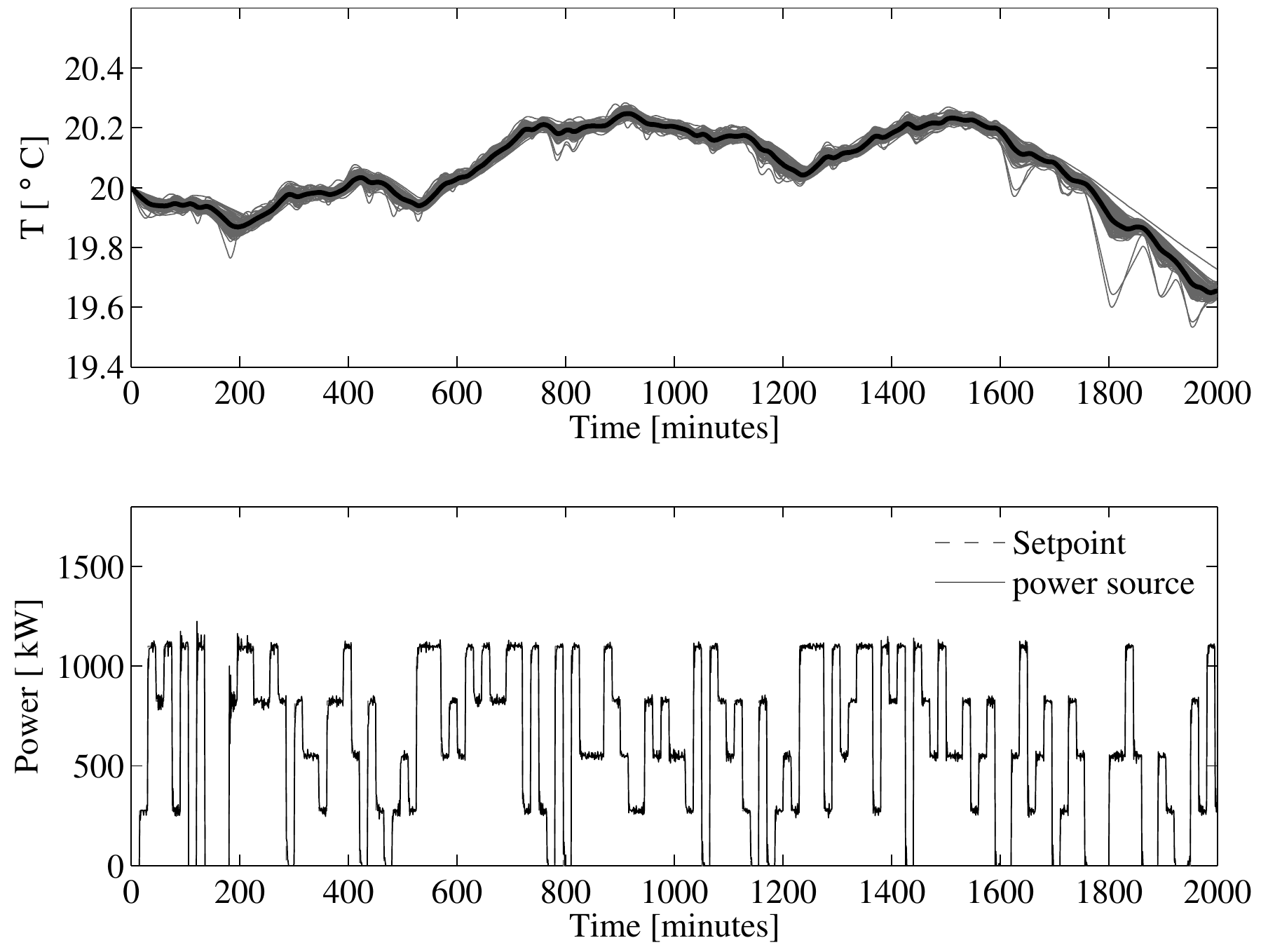}}
\caption{Top graph, the temperature dynamics of the 100 simulated buildings as is the average temperature indicated by the black line. Bottom graph, the requested thermal power at every control step indicated by the dashed line, as is the actual thermal power delivered by the CHP indicated by the black line.}
\label{fig.tracking}
\end{figure}
A first supporting result is depicted in Figure \ref{fig.tracking}, here a numerical experiment was performed where every 15 minutes a random control action was selected (within the technical specification of the CHP). The corresponding set points are depicted in the lower part of Figure \ref{fig.tracking}, as is the average thermal power as produced by the CHP. The upper graph of Figure \ref{fig.tracking} depicts the internal temperatures of the 100 buildings and the average temperature. The graph shows that when the buildings are on average in their dead band, the requested power can be tracked accurately.
Although the buildings have different physical parameters, they tend to synchronize with regard to their temperature relative to the comfort constraints. This is a direct effect of the dispatch dynamics, as those buildings with a higher priority are served first. 
 
\subsection{Energy arbitrage}
\begin{figure}[h!]
\centerline{\includegraphics[width=1\columnwidth]{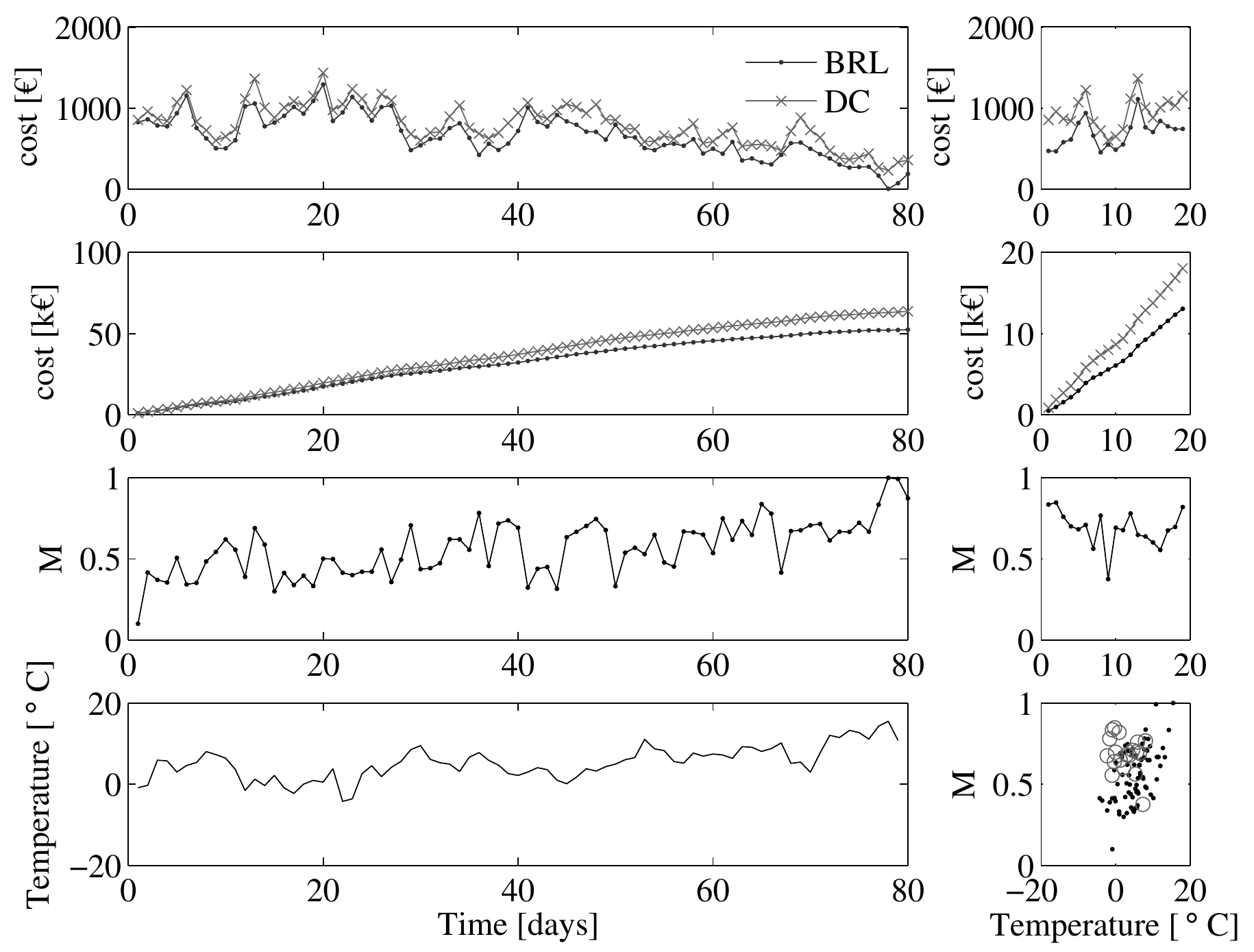}}
\caption{Overview of results obtained for an energy arbitrage scenario. The upper graphs depict the daily cost for heating for both the BRL approach and the default control case. The graphs at the second row present the cumulative cost. The graphs in the third row, depict a performance metric ($M$) as explained in the text. The lower graphs show the daily average outside temperature and the sensitivity of the performance metric relative to the average outside temperature.}
\label{fig.arbitrageOverview}
\end{figure}

In the scenario of energy arbitrage, the CHP can sell its electric energy directly at the wholesale market \cite{Kitapbayev2015823}. The energy prices are taken from the Belgian day-ahead market \cite{BelPex}, the gas price is set at 38.6 $\mbox{\euro}$/MWh \cite{eurostat}. As day-ahead prices can be predicted with a reasonable accuracy \cite{ConejoPricePrediction}, these are considered deterministic in this evaluation. Furthermore, the off line policy calculation (Algorithm 1) is performed only on a daily basis\footnote{ This is for practical reasons as the calculation of a policy typically takes 20 minutes on a Intel, 2.5 GHz, 8GB RAM}, and the simulation period covers up to 80 days.          
The results of this numerical experiment are depicted in Figure \ref{fig.arbitrageOverview}.
In the upper row, the daily cost is depicted, both for the approach presented in this paper and a default controller. The default controller applies a hysteresis controller at every building and a fixed DHN inlet temperature \cite{Dirk}. 
The cumulative cost is depicted in the second row.
It can be observed that initially the daily cost obtained with the BRL controller is similar to the cost of the default controller. 
However, as the BRL controller starts gathering more interactions, its daily cost for energy starts decreasing. To evaluate the performance during colder days, the first 20 days with a lower average outside temperature where repeated, with a policy constructed with all data. These results are depicted in the right column.
The performance is significantly improved, and the daily cost is decreased with about 20\%. Although these results are positive, they do not provide an objective metric of the quality of the control approach, since this percentage is biased by the actual price profiles. To this end, a lower bound on the daily cost is provided by taking the total energy consumed during each day and distributing this over those hours with the highest price assuming the CHP running at full capacity. This is considered an over-optimistic benchmark.
In the third row of Figure \ref{fig.arbitrageOverview} the following metric is depicted:
\begin{equation}
M = \frac{FQI-DC}{LB-DC},
\end{equation} with $DC$ the daily cost of the default controller, $LB$ the lower bound on the daily cost and $FQI$ the daily cost obtained with the solution presented in this work. A metric of 0 corresponding to the same performance as the default controller, whilst a metric of 1 corresponding to the lower bound solution. From Figure \ref{fig.arbitrageOverview} it is observed that the performance metric gradually increases to a value of around  60-70\%, also for the colder days as depicted in the right column.

\begin{figure}[h!]
\centerline{\includegraphics[width=1\columnwidth]{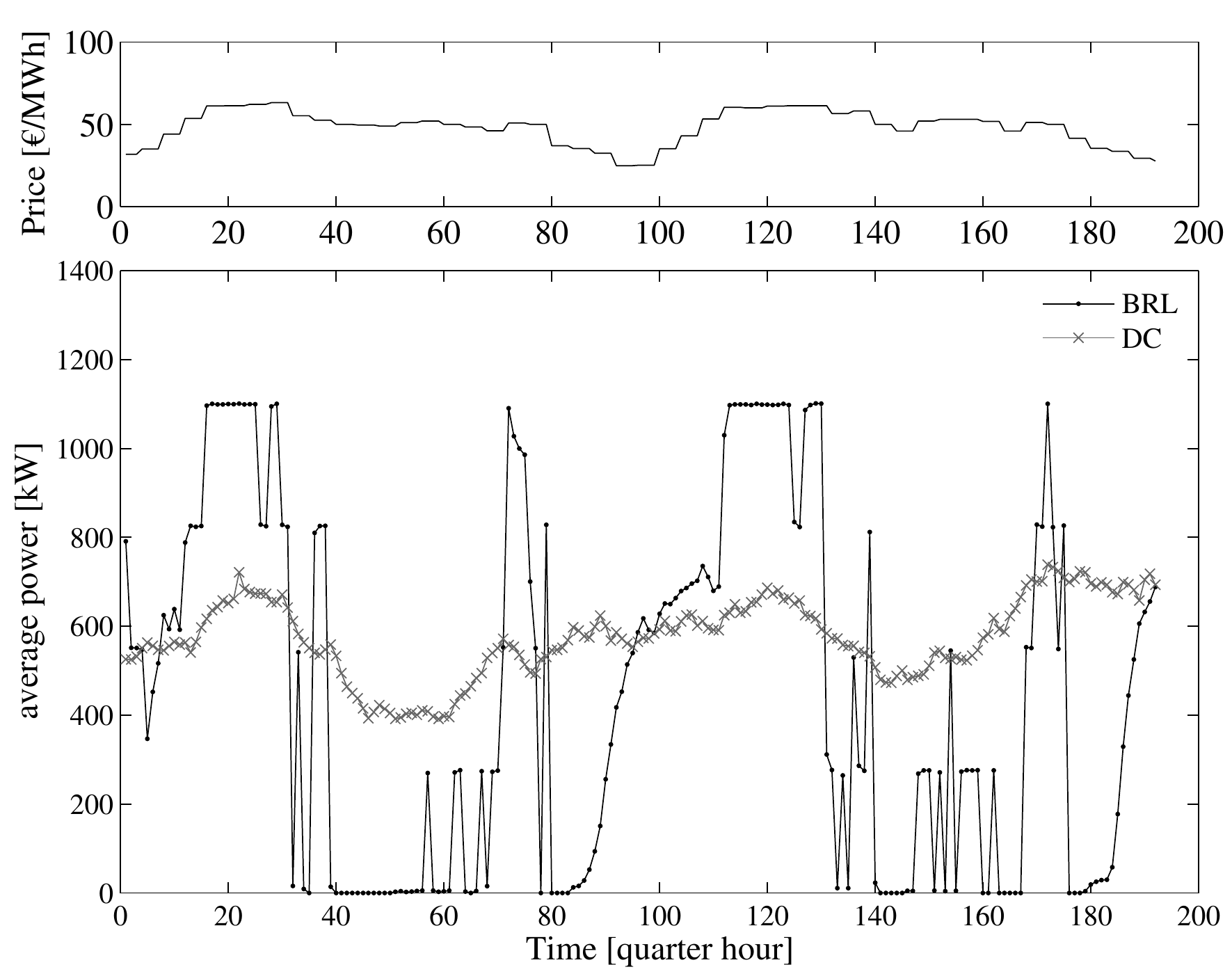}}
\caption{Top graph, the wholesale price profile.  Lower graph, the average output power, both for the default control case and the BRL approach presented in this work.}
\label{fig.powersnapshot}
\end{figure}

A snapshot of the daily power profiles of a \textit{mature} controller compared to the default controller is depicted in Figure \ref{fig.powersnapshot}. It can be seen that the controller produces heat when energy prices are high.
It is meaningful to understand that a default controller already has a reasonable performance as the heat demand is typically largest when also the wholesale price is highest. This correlation is however expected to decrease as more renewable energy comes in the production mix, making wholesale prices more volatile.


\subsection{peak shaving}
\begin{figure}[h!]
\centerline{\includegraphics[width=1\columnwidth]{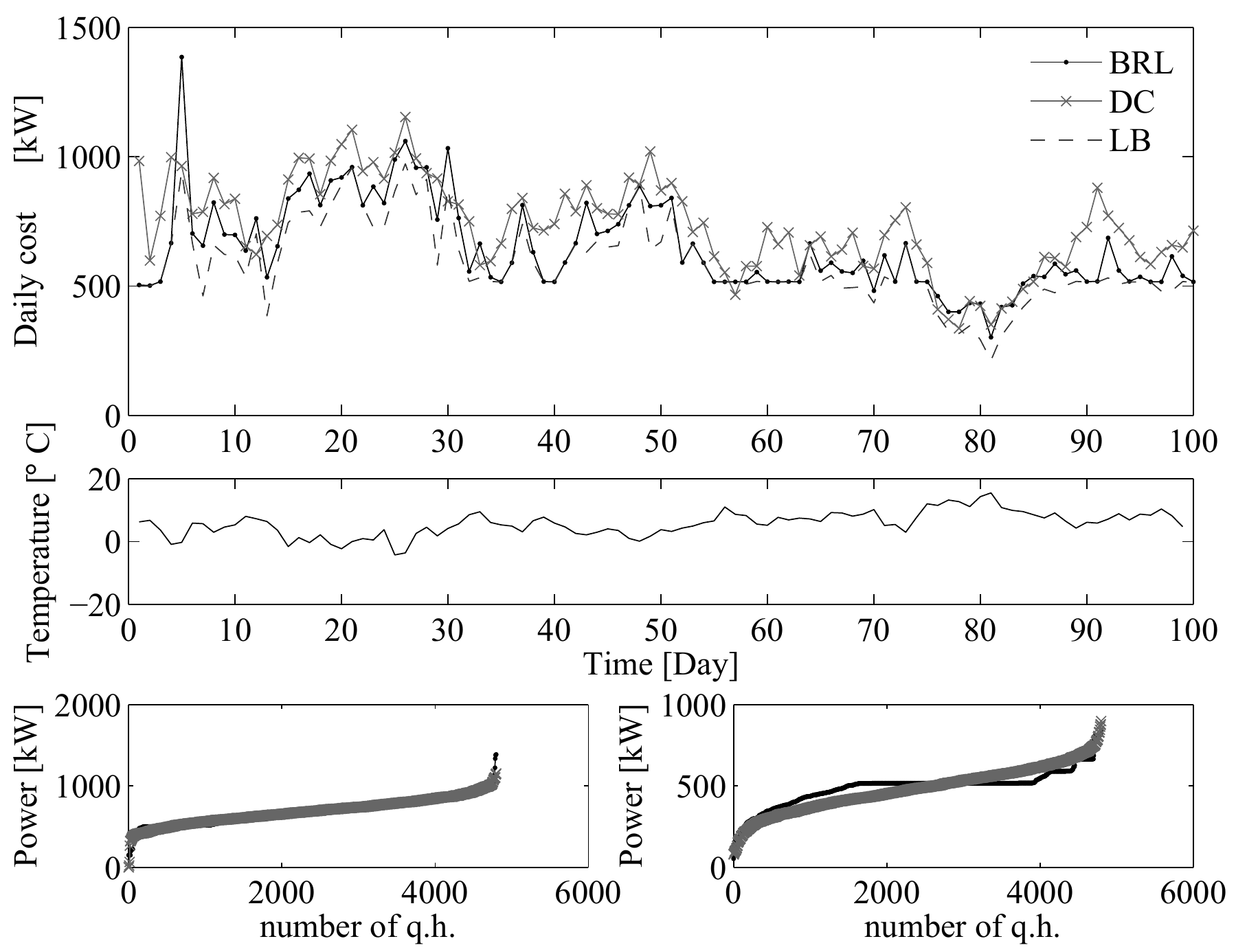}}
\caption{Top graph, daily power peak, both for the approach presented in this work and the default controller. The middle graph shows the daily average temperature. The lower row presents load duration curves for the first 50 days and for the last 50 days.}
\label{fig.PSoverview}
\end{figure}

\begin{figure}[h!]
\centerline{\includegraphics[width=1\columnwidth]{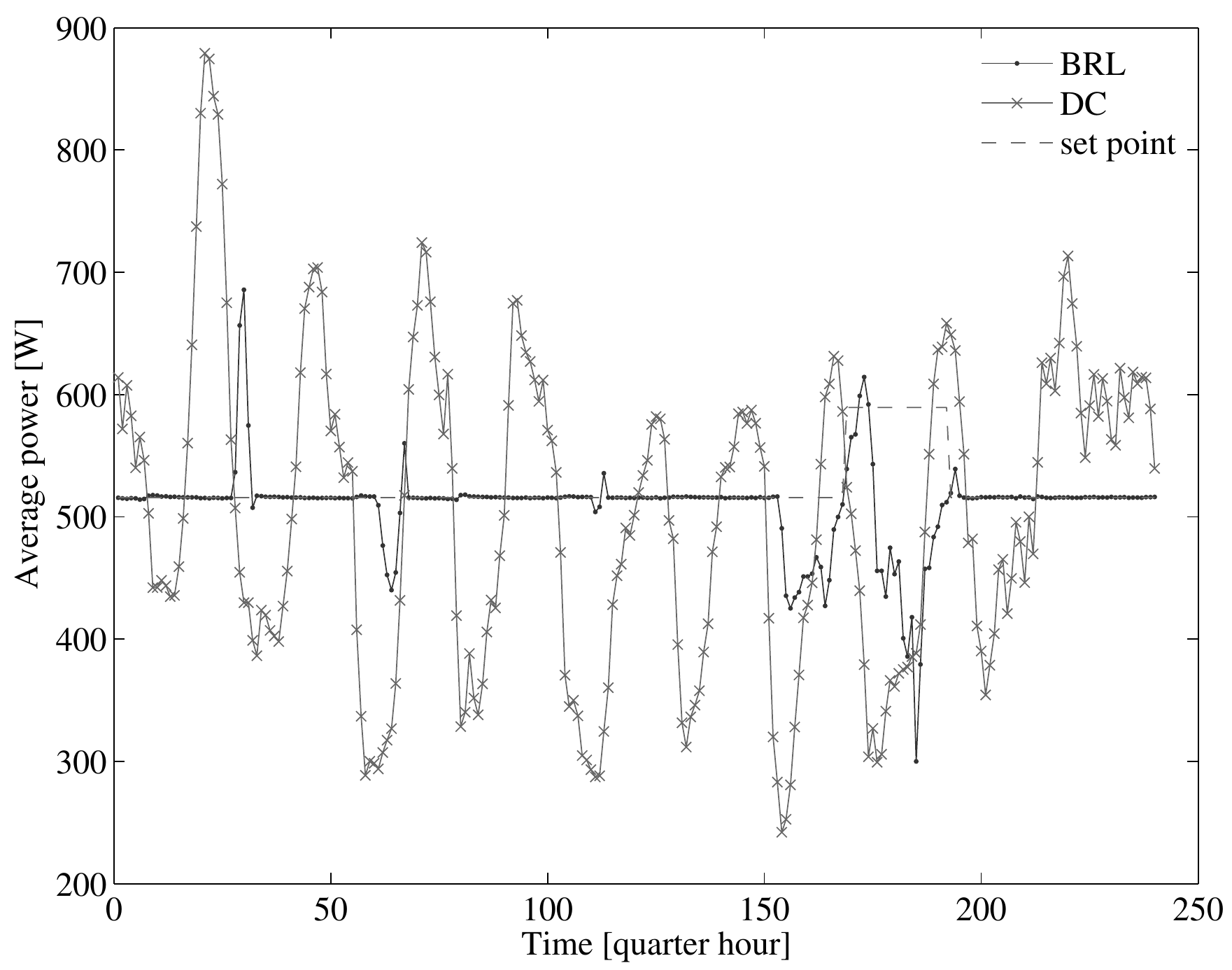}}
\caption{Power profile, both for the controller as presented in this work and a default controller in the peak shaving/valley filling scenario.}
\label{fig.PSpower}
\end{figure}

\begin{figure}[h!]
\centerline{\includegraphics[width=0.8\columnwidth]{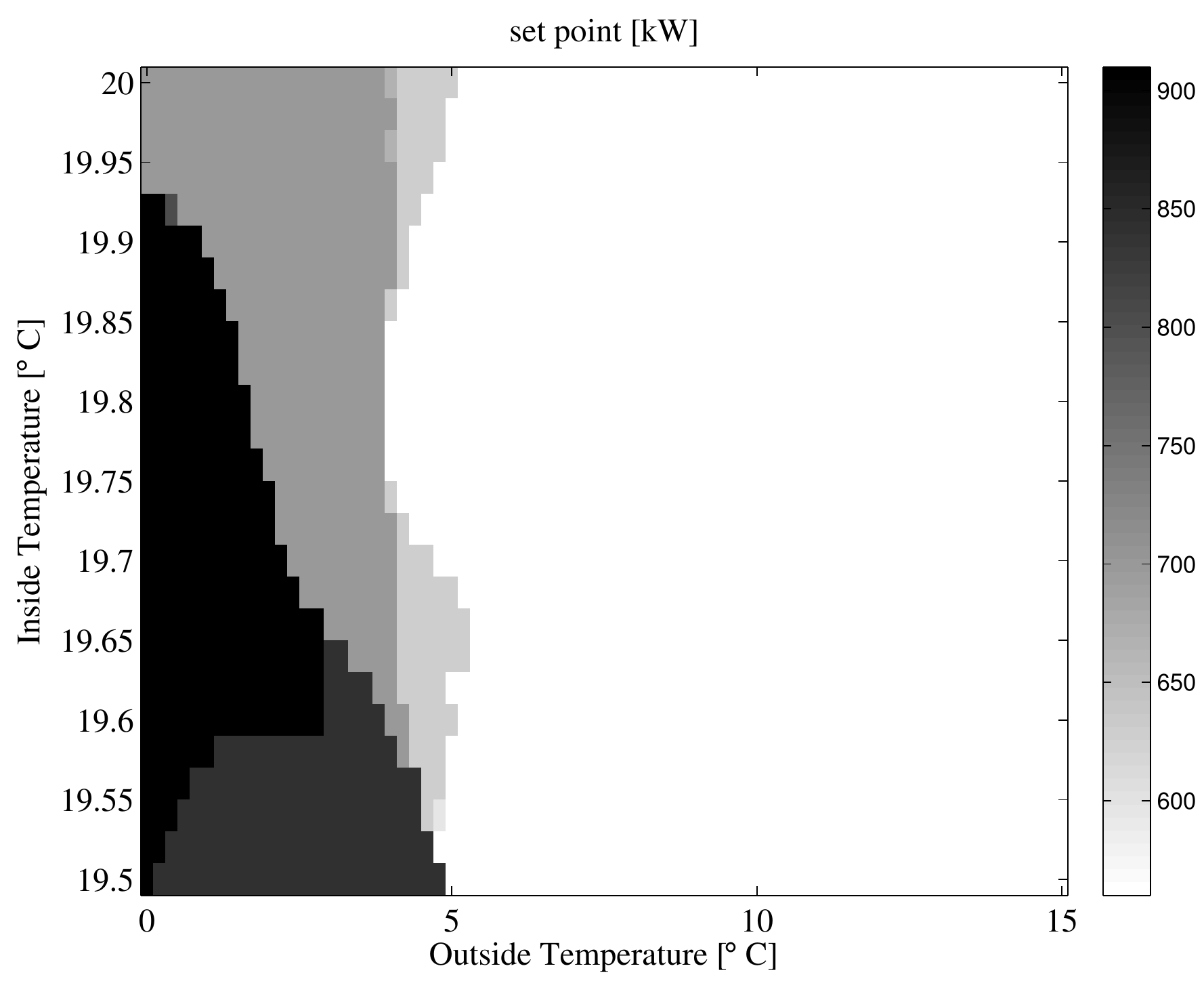}}
\caption{Policy map as obtained with the BRL approach for the peak shaving/valley filling scenario.}
\label{fig.PSpolMap}
\end{figure}

In the second experiment, peak shaving is considered. The summarizing results are depicted in Figure \ref{fig.PSoverview}. In the top graph one can see the daily peak in thermal power obtained with the default controller and the BRL approach. 
The second graph gives the average daily ambient temperature.
The daily maximum power peak indeed reduces with time. To make the performance more visible, also the load duration curves are plotted for the first and the last 50 days, both for the BRL approach and the default controller. The first 50 days there is limited improvement over the default controller, however for the subsequent period one can clearly observe the effect of the controller from the load duration curves.
A lower bound is depicted by plotting the average power corresponding to the daily energy consumed by the default control approach.
Note, on day seven (top graph in Figure~\ref{fig.PSoverview}), the profile obtained with BRL results a high peak power, which is attributed to an exploration step (\ref{eq.Boltzmann}).
The final performance is visualized more clearly in Figure \ref{fig.PSpower} where the power profile is plotted for a \textit{mature} controller. Indeed, the thermal power follows a constant profile compared to the default control case. 
A visualization of the policy obtained by the BRL controller is presented in Figure \ref{fig.PSpolMap}, here the average power set point is plotted versus the initial indoor temperature state and the expected average daily outdoor temperature. As is conceived logical, the setpoint decreases with increasing outside temperature and average air temperature.

%% file: conclusionsDHN.tex
\section{Conclusions and future work}
\label{s.conclusionsDHN}
In this work the control problem of a cluster of thermostatically controlled loads connected to a district heating network is addressed by assessing the performance of a control approach comprising a model-free reinforcement learning technique in combination with a market-based multi-agent system. The performance of the controller has been evaluated for two distinct scenarios, i.e. energy arbitrage and peak shaving. In the evaluation a detailed district heating network model has been used including hydraulic and thermal dynamics. 
For the energy arbitrage scenario, solutions are obtained that reach over 65\% of the available optimization potential after a learning period of 40 to 60 days. Knowing that the policy is updated on a daily base, this is considered a promising result. Also for the peak shaving/valley filling scenario, promising results have been obtained, since a clear performance improvement is observed.
These results support a practical implementation and \textit{coming of age} of reinforcement learning techniques. \\
\indent 
To understand the potential of completely model-free control a direct implementation of fitted Q-iteration as presented in \cite{Busoniu} has been used.
However as discussed in \cite{Ernst}, combining general domain knowledge with a model-free approach
is expected to result in an improved performance at a reduced learning time. This domain knowledge can be incorporated through e.g. information regarding the shape of the policy \cite{Busoniu} or through using a model similar as in \cite{MABRL}.
A second point of future research is directed at more automated feature extraction techniques such as autoencoders \cite{atariRL}, which is also expected to result in a reduced learning time.